\documentclass[twocolumn,floats,prb,aps,showpacs]{revtex4}

\usepackage[dvips]{graphicx}
\usepackage{amsfonts}
\usepackage{amsmath}
\usepackage{amssymb}
\usepackage{exscale}
\usepackage{eufrak}

\begin{document}

\title{The in-plane electrodynamics of the superconductivity in 
Bi$_{2}$Sr$_{2}$CaCu$_{2}$O$_{8+\delta}$: energy scales and spectral weight 
distribution}

\author{A.F. Santander-Syro}
\author{R.P.S.M. Lobo}
\author{N. Bontemps}
\email{Nicole.Bontemps@espci.fr}
\affiliation{Laboratoire de Physique du Solide, CNRS UPR 5, Ecole Sup\'erieure 
de Physique et Chimie Industrielles de la Ville de Paris, 75231 Paris cedex 5, 
France}

\author{W. Lopera}
\affiliation{Thin Film Group, Universidad del Valle, A.A. 25360, 
Cali, Colombia}

\author{D. Girat\'a}
\affiliation{Solid State Group, Physics Institute, Universidad de 
Antioquia, A.A. 1226, Medellin, Colombia}

\author{Z. Konstantinovic}
\author{Z.Z. Li}
\author{H. Raffy}
\affiliation{Laboratoire de Physique des Solides, CNRS UMR 8502, 
Universit\'e Paris-Sud 91405 Orsay cedex, France}

\date{\today}

\begin{abstract}
The {\it in-plane} infrared and visible (3~meV--3~eV) reflectivity of
Bi$_2$Sr$_2$CaCu$_2$O$_{8+\delta}$ (Bi-2212) thin films is measured between 
300~K and 10~K for different doping levels with unprecedented accuracy. The 
optical conductivity is derived through an accurate fitting procedure. We study 
the transfer of spectral weight from finite energy into the superfluid as the 
system becomes superconducting. In the over-doped regime, the superfluid 
develops at the expense of states lying below 60~meV, a conventional energy of 
the order of a few times the superconducting gap. In the underdoped regime, 
spectral weight is removed from up to 2~eV, far beyond any conventional scale.
The intraband spectral weight change  between the normal and superconducting 
state, if analyzed in terms of a change of kinetic energy is $\sim 1$~meV. 
Compared to the condensation energy, this figure addresses the issue of a 
kinetic energy driven mechanism.
\end{abstract}


\maketitle


\section{INTRODUCTION}
\label{sect:intro}

Cuprate superconductors, where the electronic correlations are important, 
deviate in many fundamental aspects from conventional superconductors. The 
electronic structure of these materials presents a strong momentum dependence, 
as demonstrated from, {\it e.g.}, angle-resolved photoemission spectroscopy 
(ARPES) experiments.\cite{Olson-FS,FSinBSCCO,Campuzano-FSinYBCO,Dessau-FS,
Campuzano-NoBilayer,Campuzano-FSinBSCO} This plays a definite role in the 
optical and transport phenomena of these high-$T_{c}$ materials,\cite{BasovVDM,
Orenstein-NodalTau} and has noticeable consequences in the pseudogap 
regime \cite{Andres-PRL-PSG} (for an experimental review of the pseudogap, see 
the reference \cite{TimuskStattPG}).

In the superconducting state, the binding of quasi-particles (QP) into Cooper 
pairs and the concomitant onset of phase coherence between the pair states are 
well understood in the case of conventional metals. The pairing mechanism and 
the formation of a coherent state remains a central problem in high-$T_c$ 
superconductivity. Various unanswered and possibly contradictory issues are 
being debated concerning the nature of the pseudogap state (PGS) and its 
temperature onset.\cite{Puchkov-PropOpt}

In this paper, we focus on a quantitative analysis in the superconducting state 
(SCS). More precisely, we investigate the spectral weight $W$, {\it i.e.} the 
area under the real part of the optical conductivity, defined as:
\begin{equation}
  W = \int_{0^+}^{\omega_c} \sigma_1(\omega,T) d\omega,
  \label{eq:DefW}
\end{equation}
where $\sigma_1(\omega,T)$ is the frequency ($\omega$) and temperature ($T$) 
dependent conductivity, and $\omega_c$ is a cut-off frequency. The optical 
conductivity being related to the absorption, the spectral weight calculated up 
to $\omega_c$ reflects the number of states available for an optical transition 
up to this energy. It has been known for a long time that there is  a noticeable 
loss of spectral weight in the optical conductivity when high $T_c$ materials 
enter the SCS. One key issue is the energy scale over which this loss of 
spectral weight occurs. It is indeed related to the excitations responsible for 
pairing. This loss of spectral weight occurs because as superconductivity sets 
in, single QP states are suppressed and the associated weight is transferred 
into the zero frequency peak associated with the superfluid condensate.

More precisely, the Ferrell-Glover-Tinkham (FGT) sum rule \cite{FGT-1,FGT-2} 
requires that the spectral weight $\Delta W$ lost when decreasing the 
temperature from the normal state into the superconducting state, must be 
retrieved in the spectral weight $W_s$ of the $\delta(\omega)$ function centered 
at zero frequency. The sum rule is exact if integrating up to infinity, and 
results merely from charge conservation. Actually, the sum rule is satisfied 
($\Delta W \simeq W_s$) provided the integration is performed up to a large 
enough value $\hbar\Omega_M$. In conventional superconductors, $\hbar \Omega_M$ 
is typically $16 k_{B}T_c$, or about $4 \Delta$  ($\Delta$ is the 
superconducting gap).\cite{FGT-1,FGT-2} $\hbar \Omega_M$ is considered to be a 
characteristic energy of the boson spectrum responsible for the pairing 
mechanism. Would cuprates display this conventional behavior, then taking a 
typical maximum gap magnitude (in a d-wave superconductor) $\Delta_M$ of 25 meV 
yields $\hbar \Omega_M \sim  0.1$~eV.\cite{GapHTc-1,GapHTc-2} Indeed, from 
conventional BCS theory, including strong electron-phonon coupling, the amount 
of violation of the FGT sum rule is of the order of $(\Delta/\omega)^2$ for 
$\omega \gg 2\Delta$, hence negligible.\cite{Maksimov-SumRule}

An apparent violation of the sum rule, i.e. $\Delta W < W_s$ when integrating
up to 0.1 eV was observed from interlayer conductivity data: exhausting the sum 
rule could then require an anomalously large energy scale, which was suggested 
to be related to a change of interlayer kinetic energy when the superfluid 
builds up.\cite{Basov,KatzBasov,BasovVDM} Although early measurements yielded 
conventional  {\it in-plane} energy scales \cite{Basov}, the decrease of the 
{\it in-plane} kinetic energy, suggested long ago as a possible pairing 
mechanism in the framework of the hole undressing 
scenario,\cite{Hirsch1-1,Hirsch1-2} was given renewed 
interest.\cite{Hirsch3-1,Hirsch3-2,Hirsch2} Indeed, later ellipsometric 
measurements and IR reflectivity 
experiments\cite{Rubhausen-100Delta,vdmScience,Boris} showed that in-plane 
spectral weight was lost up to the visible range. The key issue which has to be 
addressed experimentally is whether this spectral weight is transferred into the
condensate.\cite{Andres-EPL}

The data presented in this paper report the investigation of the FGT sum rule in 
three carefully selected thin films from the Bi-2212 family, probing three 
typical locations in the phase diagram: the underdoped (UND), the optimally 
doped (OPT) and the overdoped (OVR) regime. A detailed study  allows to work out 
with well controlled error bars the FGT sum rule and the spectral weight 
changes. We find that within these error bars, retrieving the condensate 
spectral weight in the OVR and OPT samples requires integrating up to an energy 
of the order of 0.1~eV (800 cm$^{-1}$), {\it i.e.} a conventional energy scale. 
In the UND sample, about 20\% of the FGT sum rule is still missing at 1~eV, and
the integration must be performed up to at least 16000 ${\rm cm}^{-1}$ (2~eV), 
an energy scale much larger than typical boson energies in a solid, and 
$\sim 100$ times larger than the maximum gap. We derive the associated change of 
the in-plane kinetic energy, which turns out to agree with some theoretical 
calculations.\cite{Hirsch3-1,Hirsch3-2,NormanPepin-QPFormation} This paper 
supplements our previous report on the FGT sum rule,\cite{Andres-EPL} by 
developing in detail the investigation on the spectral weight distribution in 
the superconducting state, and by presenting the details of the analysis of the 
thin-film reflectivity and of the uncertainties.

The paper is divided as follows: after a presentation of the experimental 
results, we will perform the analysis of the data in terms of the partial 
spectral weight and the  FGT sum rule as a function of the cut-off frequency. 
We will give an interpretation of the results in terms of a change of the 
in-plane kinetic energy. In the appendix, we describe the sample fabrication, 
characterization and selection, and the experimental set-up. Then, we discuss 
the procedure to derive the optical functions from the reflectivity of a film 
deposited onto a substrate, giving a detailed account of the experimental 
uncertainties arising from this procedure, and explaining how we incorporated 
the uncertainties associated with the low temperature extrapolation of the 
normal-state spectral weight (which is required to compute correctly the FGT sum 
rule) into the total error bars.


\section{EXPERIMENTAL RESULTS}
\label{sec:results}

\subsection{Samples and measurements}

\begin{figure*}
  \begin{center}
    \begin{tabular}{lr}
      \includegraphics[height=10cm]{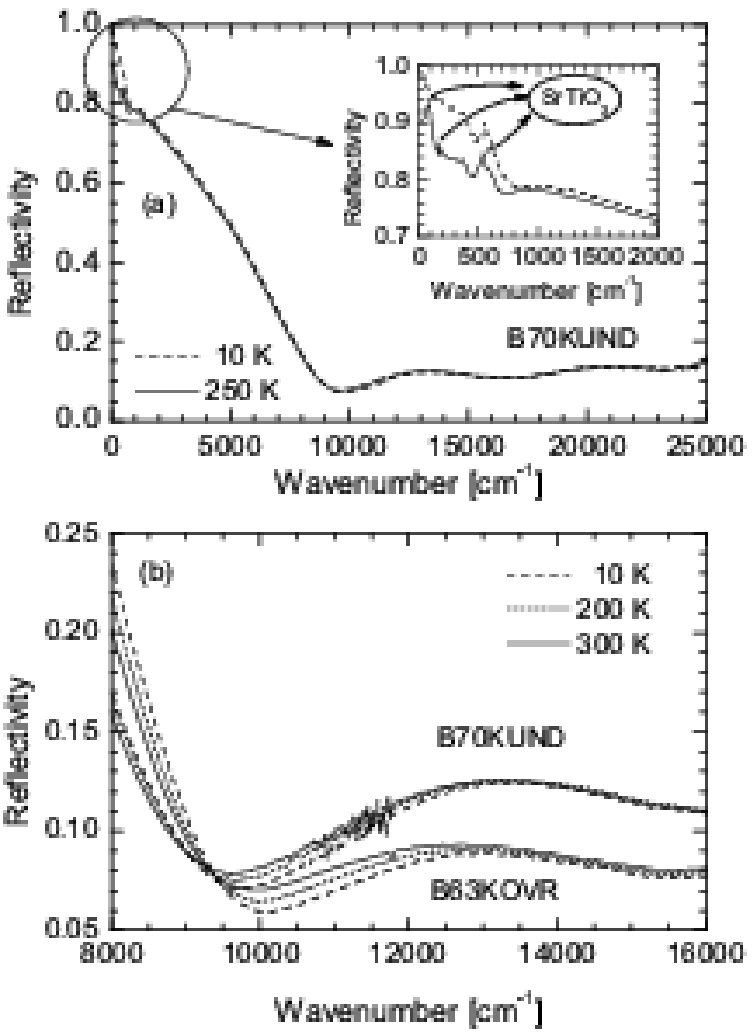} &
      \includegraphics[height=10cm]{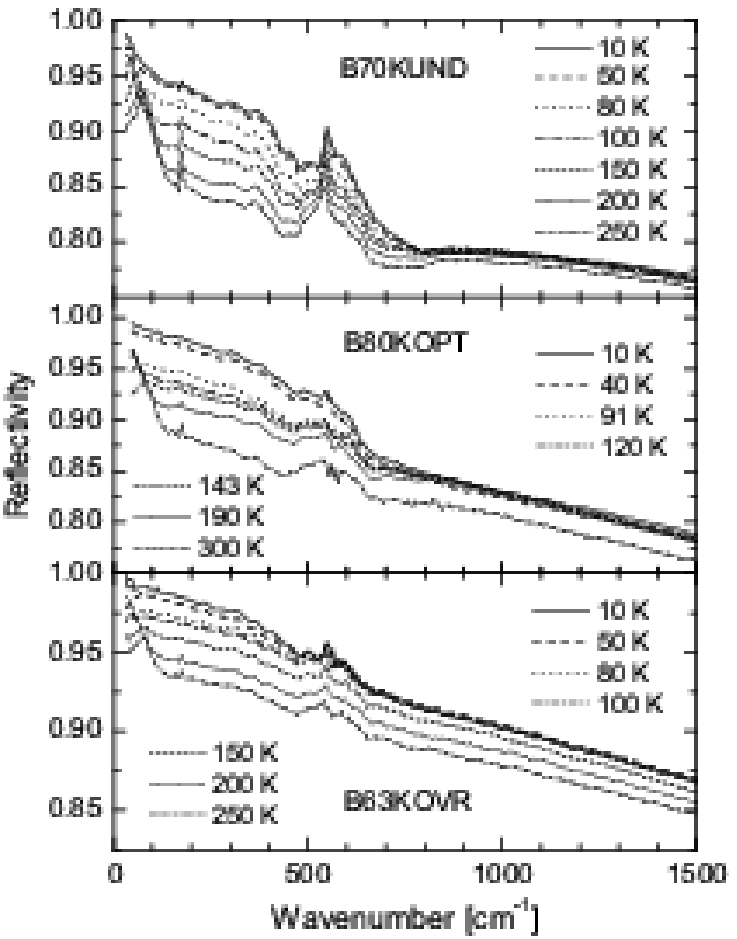}
    \end{tabular}
  \end{center}
  \caption{\label{Fig1}
    (a) Reflectivity of the B70KUND sample in the
    whole experimental range, at high (250~K) and low (10~K) temperatures.
    The inset shows the variation of the reflectivity in the far-infrared,
    and the phonon peaks from the substrate. (b) Temperature
    changes of reflectivity in the visible range for the B70KUND and B63KOVR
    samples. Right panels: reflectivity of the three samples, for a
    restricted set of temperatures, up to 1500~cm$^{-1}$.}
\end{figure*}

The thin films studied in this work were epitaxially grown by r.f. magnetron 
sputtering on (100) SrTiO$_3$ substrates heated at temperatures 
$\gtrsim 700^\circ$C.\cite{Li-Films,Zorica-Films} X-ray analyses confirmed that 
the films are single phase, with the {\it c}-axis perpendicular to the 
substrate, and a cation ratio (in particular a Bi$/$Sr ratio) corresponding to 
the stoichiometric one.\cite{Zorica-Thesis} Their maximum critical temperature 
(defined at zero resistance) obtained in these conditions is $\sim 84$~K. The 
$\mathbf{a}$ and $\mathbf{b}$ axes show a single orientation (45$^{\circ}$ with
respect to the substrate axes), with possible exchanges between $\mathbf{a}$ and 
$\mathbf{b}$ from one grain to another. The directions -Cu-O-Cu- are parallel to 
the (100) and (010) axes of the substrate. Films prepared by the above procedure 
are usually in a nearly optimally doped state. The various doping levels (UND 
and OVR) were obtained by post-annealing the films in a controlled 
atmosphere.\cite{Zorica-Films} In total, thirteen samples of Bi-2212 thin films 
were investigated. All the films were first characterized by electrical 
resistance measurements, and a first estimate of their thicknesses was obtained 
by Rutherford Back-Scattering (RBS) on samples prepared under exactly the same 
conditions as the ones used in this work, on MgO substrates. Absolute 
{\it resistivities} were not measured, as this would have required a 
lithographic etching of the samples, which is incompatible with the optical 
measurements. As a next step, their optical homogeneity in the mid-infrared was 
characterized by infrared microscopy ($\mu$IR), as will be described in the 
appendix. The surface quality in the visible was characterized by optical 
microscopy. 

Another Bi-2212 film on (100) LaAlO$_3$ single crystal substrate, prepared by a 
high-pressure dc sputtering technique, was also studied.~\cite{Prieto} In this 
case, pure oxygen at a 3.5 mbar pressure was used as sputtering gas, with a 
880$^\circ$C deposition temperature. Typical deposition rates being 1000~\AA/h, 
the thickness of Bi-2212 films lies around 4000~\AA. Films were post-annealed 
for 45 min at 0.01 mbar oxygen pressure, yielding films with an optimal oxygen 
content (maximum critical temperature $T_c$=90~K). Such fabricated films are 
single phase and c-axis oriented, as confirmed by different techniques. In 
$\theta - 2 \theta$ scan X-ray measurements, only (001) peaks were observed. The 
spread of the distribution of c-axis oriented grains tilted away from the 
surface was obtained from the rocking curves. Such rocking curves around the 
(00\u{10}) diffraction peak display a width at half-maximum~$0.2-0.3^\circ$, 
indicating good crystalline quality of the samples. RBS combined with channeling 
measurements allowed to verify the composition and epitaxial quality of the 
layers. A mean roughness of around 4.5 nm in the surface of the films was found 
by atomic force microscopy (AFM). Because of their large thickness, the 
substrate contribution in such samples is almost negligible, providing us with a 
valuable test of the comparison between Kramers-Kronig data processing and our 
customized fitting procedure, as explained below. Finally a similar 
YBa$_2$Cu$_3$O$_7$ (YBCO) film at optimal doping, deposited onto LaAlO$_3$ was 
also studied for comparison (see appendix).

The full infrared-visible reflectivities, taken at typically 15 temperatures 
between 10~K and 300~K and at quasi-normal incidence ($\sim 8^{\circ}$), were
measured for all the films in the spectral range $[30-7000]$~cm$^{-1}$ with a 
Bruker IFS 66v Fourier Transform spectrometer, supplemented with standard 
grating spectroscopy in the range $[4000-28000]$~cm$^{-1}$ (Cary-5).

To determine the absolute value of the reflectivity, we used as unity
reflectivity references a gold mirror in the $30-7000$~cm$^{-1}$ spectral range, 
and a silver mirror in the remaining spectral range. To ensure accuracy in this 
absolute measurement, the sample holder was designed to allow to commute between 
the reference and the sample, placing one or the other at the same place, within 
an angular accuracy better than $10^{-3}$~rad, on the optical path, 
independently of the optical (light-entrance windows) or thermal (temperature) 
set-up of the cryostat. The latter is a home-built helium-flow cryostat, with
the sample holder immersed in the helium flow, so that both sample and reference 
mirror are cooled simultaneously. The temperature in our set-up can be 
stabilized within 0.2~K. A circular aperture placed in front of the 
sample/reference position guarantees that the same flux is irradiating the 
sample and the reference. The temperature of the sample/reference block is 
measured independently of that of the helium flow. Once thermal equilibrium has 
been achieved in the entire cryogenic set-up, consecutive measurements of the 
sample and reference spectra can be done, under exactly the same conditions. 
Our accuracy in the determination of the {\it absolute} reflectivity is 1\%.

In order to extract the optical functions from our raw measurements, we need to 
know the film thicknesses (as explained below). After the infrared measurements, 
the film thicknesses were directly determined by RBS for the OPT and OVR 
samples, yielding 395 nm and 270 nm respectively (The uncertainty in the 
thickness given by RBS on films grown on SrTiO$_3$ is 300~{\AA}). A lower bound 
for the UND sample was estimated by RBS in a sample grown on MgO simultaneously 
to ours, yielding 220 nm. (The UND sample itself was destroyed in an 
unsuccessful attempt to measure its thickness by transmission electron 
microscopy).

It is known that temperature changes of the optical response of cuprates in the 
mid-infrared and the visible ranges are small, but, as discussed further, cannot 
be neglected.\cite{Maksimov} Yet, as remarked by van der Marel and 
coworkers,\cite{vdMTrieste} most studies rely on a single spectrum at one 
temperature in the visible. Moreover, few temperatures are measured, due to the 
lack of resolution when the temperature-induced reflectivity changes are small. 
Using thin films rather than single crystals allowed us to measure 
{\it relative} variations in the reflectivity within less than 0.2\%, even in 
the visible range, due to their large surface (typically $6\times 6$~mm$^{2}$). 
We were thus able to monitor the temperature evolution of the reflectivity 
spectra in the full available range (30-28000~cm$^{-1}$). This is obviously 
important if one is looking for a spectral weight transfer originating from (or 
going to) any part of the {\it whole} frequency range.

\subsection{Raw data}

Our films critical temperatures are 70~K (B70KUND), 80~K (B80KOPT) and 63~K 
(B63KOVD). Figure~\ref{Fig1}(a) shows two reflectivity curves (at $T=250$~K and 
$T=10$~K) of the B70KUND sample. This is an example of the typical spectra we 
have measured. The signal-to-noise ratio is unprecedented, and relative 
variations of the reflectivity of the order of 0.2\% can be measured. In the 
far-infrared (see the inset), the three phonon peaks coming from the underlying 
substrate are clearly visible, specially at high temperatures (or low doping), 
when the system is less metallic. The lowest-energy peak is the soft mode of 
SrTiO$_3$. We experimentally determined the optical constants of the SrTiO$_{3}$ 
at each temperature, so as to take into account the changes in reflectivity due 
to changes in the substrate optical properties alone. This is important when 
extracting the intrinsic optical constants of the Bi-2212 (the procedure will be 
described below).

\begin{figure*}
  \begin{center}
    \begin{tabular}{lr}
      \includegraphics[width=8cm]{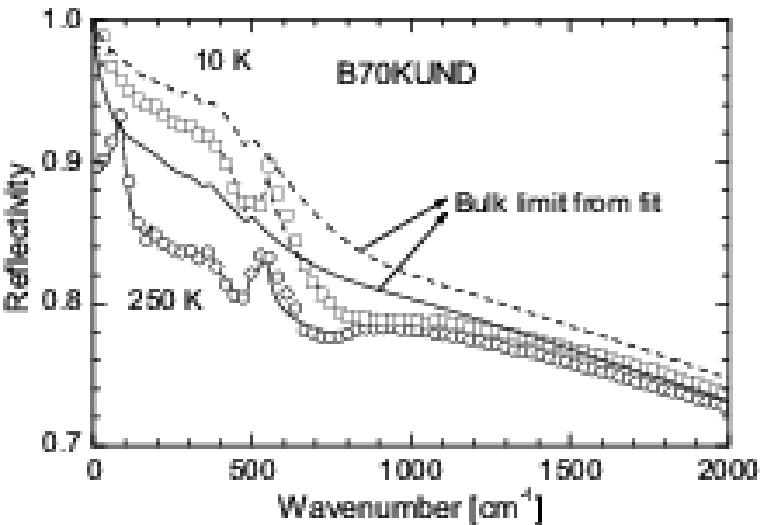} &
      \includegraphics[width=8cm]{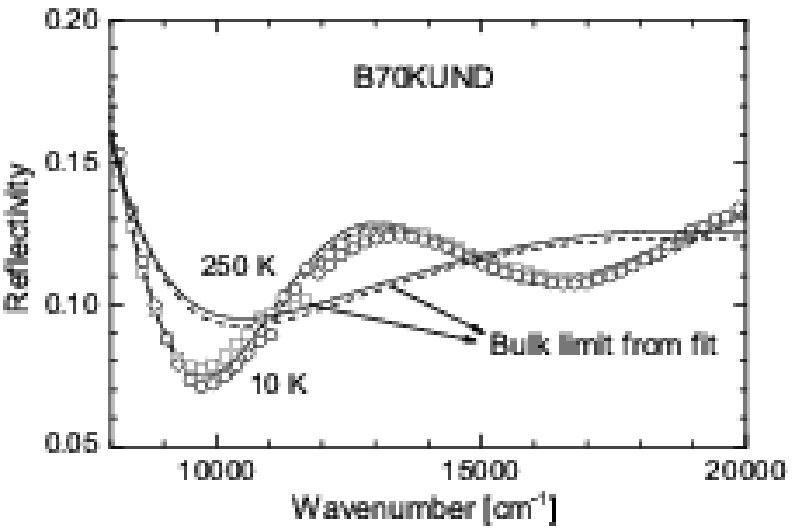}
    \end{tabular}
  \end{center}
  \caption{ \label{Fig2}
    Measured reflectivity spectra of the B70KUND sample at 250~K and
    10~K (open symbols) and fitted spectra (lines) at far-infrared (left
    panel) and visible (right panel) frequencies.  The fits lie within
    $\pm 0.1$\% of the raw data in the full spectral range
    (up to 25000~cm$^{-1}$).
    The bulk reflectivities obtained from the fits
    are also shown.  Note that
    the contribution from the substrate to the raw data is
    effectively suppressed in the bulk reflectivities.
   }
\end{figure*}

While the major temperature changes in the reflectivity occur at low frequency, 
where reflectivity increases with increasing doping or decreasing $T$, changes 
up to the visible range (up to $\sim 12000-14000$~cm$^{-1}$) are important as 
well. This is illustrated for the B70KUND and B63KOVR samples in 
Fig.~\ref{Fig1}(b). This figure shows indeed that, in the visible range, a 
decrease in doping has the same qualitative effect that increasing the 
temperature: in both cases, the reflectivity in the visible range increases. 
This effect disappears below temperatures close to $T_{c}$:  there is no 
measurable change in the visible reflectivity below $\sim 100$~K. These opposite 
temperature behaviors of the infrared and visible reflectivities have been 
observed as well by temperature-modulated differential spectroscopy 
measurements.\cite{Holcomb,Little} The oscillations in the reflectivity in the 
visible range come from both interband transitions and interferences of the 
light bouncing back and forth inside the film. The period of these oscillations, 
and the shape of the reflectivity in the vicinity of the plasma edge, depend on 
the sample thickness.

The right panels of Fig.~\ref{Fig1} show the reflectivities of the three 
samples, for a restricted set of temperatures, up to 1500~cm$^{-1}$. Note that, 
upon increasing doping, the reflectivity increases for a given temperature in 
this spectral range, as a consequence of the material becoming more metallic.

\subsection{Extraction of the optical functions}

The contributions of the substrate to the measured reflectivities preclude the 
Kramers-Kronig (KK) analysis on thin films. In order to extract the optical 
functions intrinsic to Bi-2212, we simulated its dielectric function at each 
temperature and doping levels using Drude-Lorentz oscillators (thus warranting 
causality). In the superconducting state, where a superfluid exists, we used a
London oscillator as well. The London oscillator is indeed necessary to simulate 
properly the change in slope in the reflectivity observed at low frequencies and 
low temperatures (Fig.~\ref{Fig1}, right panels). We then modelled the 
reflectance of the film on top of a substrate, using the optical constants of
SrTiO$_3$ that, as already stated, were experimentally determined for each 
temperature. We found that the best description of our data was obtained by 
assuming an infinitely thick substrate ({\it i.e.}, no backward reflection from 
the rear of the substrate; see the Appendix for details). Finally, we adjusted 
the attempt dielectric function of the Bi-2212 in order to fit accurately the 
raw reflectivity spectra. Examples of such fits, for the B70KUND sample, are 
shown in Fig.~\ref{Fig2} at far-infrared (left panel) and visible (right panel) 
frequencies. Once the dielectric function is known, we can generate any other 
optical function, in particular the optical conductivity.

We have verified that, in average over the whole experimental range, a relative 
error $\Delta R / R$ in the fit yields a relative error magnified by at most a
factor 10 in the real part of the optical conductivity, provided that 
$\Delta R / R \ll 1$. These relative errors spread over $2-3$ times the range 
where the deviation to the raw data occurs (see the Appendix, in particular
Fig.~\ref{Fig9}, for details). The thicknesses of the films are also determined 
by the fit. Fits are accurate within less than 0.5\% taking 241, 434 and 297~nm 
for the B70KUND, B80KOPT and B63KOVR samples respectively. These values differ 
by less than 10\% from the RBS measurements. We checked that the associate error 
in the conductivity, within the experimentally measured range, is less than 
10\%. The fit yields a valuable extrapolation of the conductivity in the 
low-energy range ($\hbar \omega < 30$~cm$^{-1}$, not available 
experimentally),\cite{Quijada} which is important in the evaluation of the 
spectral weight. In this range, however, the relative error in the conductivity 
was calculated and reaches 20\%.\cite{Andres-PRL-PSG}

We have used the thick optimally doped film ($T_{c}$=90~K) as a further check of 
the validity of the fitting procedure. We have compared the conductivities 
extracted by the fitting procedure (after full characterization of the LaAlO$_3$ 
substrate) to the ones obtained by performing a Kramers-Kronig transform of the 
raw reflectivity data. The reason for this attempt is that the fits show a 
reconstructed bulk reflectivity for this sample differing within less than 1~\% 
from the raw spectrum, in the whole spectral range. This turns out to have a 
dramatic impact on the conductivity below 100~cm$^{-1}$. However, above this 
frequency, the conductivities deduced from the fit and from KK transform lie 
within 20~\% one from the other, with a systematic trend placing the KK result 
above the result from the fit. This observation is entirely compatible with all 
the estimates presented in the appendix. It shows moreover that the fitting is 
absolutely required  for the low frequency analysis and the eventual spectral 
weight calculations: it is the low frequency range where the penetration length
of the electromagnetic wave is large and therefore where the substrate effects 
are the most detrimental. 

\begin{figure}
  \begin{center}
      \includegraphics[width=7cm]{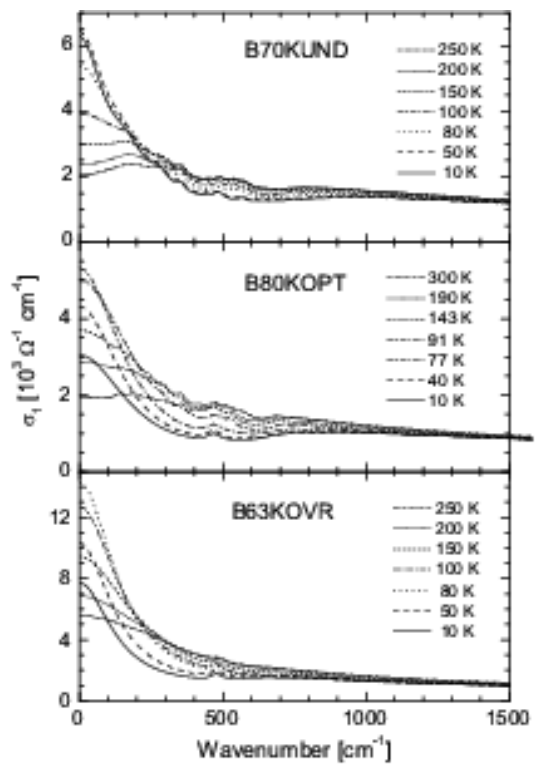}
  \end{center}
  \caption{\label{Fig3}
    Selection of conductivity spectra for the B70KUND (upper), B80KOPT (middle) 
    and B63KOVR (lower) samples. For the B70KUND sample, within the showed
    spectral range, the spectra at 50~K and 10~K are indistinguishable.
    The error bars in the conductivity spectra are
    $\Delta \sigma / \sigma \lesssim 10$\% for
    $\hbar \omega > 30$~cm$^{-1}$, and $\Delta \sigma / \sigma\sim 20$\%
    for $\hbar \omega < 30$~cm$^{-1}$.}
\end{figure}

\begin{figure}
  \begin{center}
    \includegraphics[width=7cm]{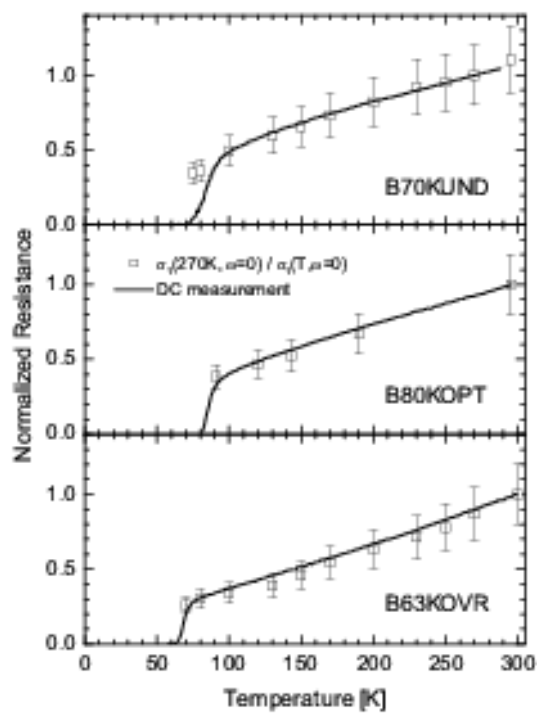}
  \end{center}
  \caption{\label{Fig4}
    Measured resistance of the three selected samples (normalized to unity at
    high temperatures), and comparison with the DC extrapolation of the
    respective conductivities obtained from the reflectivity spectra.}
\end{figure}

Figure~\ref{Fig3} shows the conductivity spectra for the B70KUND, B80KOPT and 
B63KOVR samples, for the same temperatures and in the same spectral range as the
reflectivity spectra of Fig.~\ref{Fig1}.  Note that the conductivity values are 
larger for the overdoped sample, in agreement with a larger density of charge 
carriers. The low-frequency values of the conductivity are reasonable figures 
for Bi-2212: at 250~K, for example, our DC extrapolations give as resistivities 
$\rho_{ab} \approx 500$~$\mu \Omega$-cm for the underdoped sample, and 
$\rho_{ab} \approx 160$~$\mu \Omega$-cm for the overdoped one. Furthermore, as 
can be seen in Fig.~\ref{Fig4}, the temperature behavior of the DC 
extrapolations is in agreement with the resistance measurements. For the 
underdoped sample, the two data points within the resistive transition do not 
follow the trend of the measured resistance.  Whatever the phenomenon causing 
the large resistive transition (a well documented feature in underdoped Bi-2212
films~\cite{Eckstein-Raffy}) it is not taken into account in our analysis,
which assumes an homogeneous material. In this respect, the London oscillator 
used for our fits was only ``turned-on" at $T<T_{c}$ (defined by 
zero-resistance).

In the normal state, the conductivity spectra of the three samples are 
qualitatively similar.  The weak shoulder at $\sim 200$~cm$^{-1}$ visible in the 
UND sample down to 150~K has been observed in single crystals as 
well,\cite{Puchkov-PropOpt} and remains unexplained so far. The low-frequency
conductivity ($\hbar \omega \lesssim 200$~cm$^{-1}$) of the three samples 
increases when the temperature decreases (metallic behavior), while at higher 
frequencies, because of the conservation of the spectral weight, the 
conductivity decreases when the temperature drops. The conservation of the 
spectral weight in the normal state is retrieved upon integration up to 
$\sim 2$~eV.  The implications of such observation on the electrodynamics of the 
normal state have already been analyzed.\cite{Andres-PRL-PSG}

The superconducting transition is marked, for the overdoped and optimally-doped
samples, by a decrease of the conductivity over the spectral range shown in 
Fig.~\ref{Fig3}. This effect becomes more pronounced as the temperature 
continues to drop, so that a clear loss of spectral weight, associated with the 
formation of the zero-frequency condensate, is observed in this spectral range. 
In contrast, the low-frequency ($\hbar \omega \lesssim 100$~cm$^{-1}$) 
conductivity of the underdoped sample does not decrease when temperature 
decreases below $T_{c}$, and a large Drude-like contribution persists in the 
superconducting state. Beyond this energy scale, the normal- and 
superconducting-state conductivities cross, and there is no clear loss of 
spectral weight within the spectral range shown in the figure.  A deeper 
analysis is needed in this case, based on the FGT sum rule.


\section{ANALYSIS AND INTERPRETATION}
\label{sec:analysis}

\subsection{In-plane FGT sum rule for Bi-2212}

Be $T_{A} \geq T_{c}$, and $T_{B} < T_{c}$.  From an experimental point of view,
the FGT sum rule compares the change in spectral weight
$\Delta W = W(T_{A})-W(T_{B})$ (Eq.~\ref{eq:DefW}) and the supefluid weight 
$W_{s}$.

Within the London approximation, at frequencies below $\Delta_M$, the slope of 
the real part of the dielectric function $\epsilon_{1}\left(\omega \right)$ 
plotted versus $1/\omega^{2}$ should be directly related (as was done in our 
previous report~\cite{Andres-EPL}) to the superfluid spectral weight $W_{s}$ 
through the ``London'' frequency $\Omega_{L}=c/\lambda_{L}$, where $\lambda_{L}$ 
is the London penetration depth. However, for the underdoped sample, the 
presence of a large Drude-like contribution to the conductivity in the 
superconducting state yields, by this procedure, an overestimate of about 
$15-20$\% on $\Omega_{L}^{2}$ ({\it i.e.}, on the superfluid 
weight).\cite{MN-Private} A simulation of this effect shows that, for this 
sample, the best estimate for the value of $\Omega_{L}$ is the input parameter 
of the fit, whose accuracy (within $0.1\%$) we have carefully re-worked out. For 
the optimally doped and overdoped samples there is no significant difference 
between the slope of the $\epsilon_{1}$ versus $\omega^{-2}$ plot and the 
superfluid input parameters of the fits, in accordance with a weaker 
contribution of the charge carriers to the superconducting low-frequency 
conductivity (Fig.~\ref{Fig3}, middle and lower panels).

In this way we find that, at 10~K, $\lambda_{L}=6250$~{\AA}, 2900~{\AA} and 
2250~{\AA} for the B70KUND, B80KOPT and B63KOVR samples respectively. The values 
for the overdoped and optimally doped samples are in fair agreement with those 
reported in the literature.\cite{Villard-Lambda} There are no reliable data on 
the absolute value of the London penetration depth for underdoped 
samples.\cite{DiCastro-Lambda}

\begin{figure}
  \begin{center}
    \includegraphics[width=8cm]{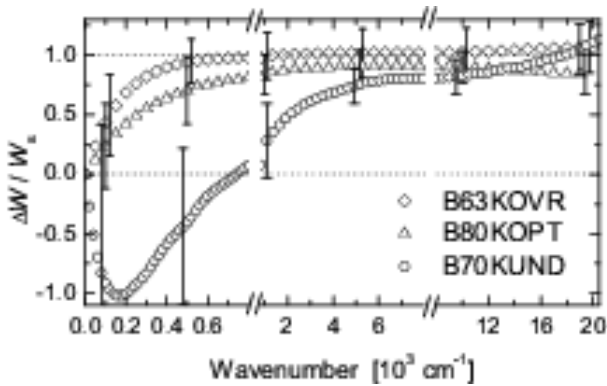}
  \end{center}
  \caption{\label{Fig5}
         Ratio $\Delta W / W_s$ versus frequency showing the exhaustion
         of the FGT sum rule at conventional energies for the OVR (solid lines,
         right error bars) and OPT (dashed lines, middle error bars) samples.
         An unconventional ($\sim 16000~{\rm cm}^{-1}$ or 2 eV) energy scale
         is required for the UND sample (short-dashed lines, left error bars). 
         Note that the frequency scale changes at $800$ and $8000$~cm$^{-1}$. 
         The changes in spectral weight are taken between 80~K--10~K, 91~K--10~K
         and 100~K--10~K for the OVR, OPT and UND samples, respectively.
         }
\end{figure}

Figure~\ref{Fig5} shows the ratio $\Delta W / W_{s}$ for the three samples 
reported in this work. The changes in spectral weight are taken between 
80~K--10~K, 91~K--10~K and 100~K--10~K for the OVR, OPT and UND samples 
respectively, so that the normal-state temperature lies slightly above the 
resistive transition regime. In the underdoped sample, we find that at energies 
as large as 1~eV (8000~cm$^{-1}$) $\Delta W / W_{s} \sim 0.8 \pm 0.15$ (details 
about the evaluation of the error bars will be given below). It approaches 1 
at $\sim 16000$~cm$^{-1}$. A Kramers-Kronig based analysis shows that, beyond 
this frequency range, the error in the determination of the conductivity (hence 
the spectral weight) rapidly increases.  This is a consequence of the 
reflectivity for $\hbar \omega > 25000$~cm$^{-1}$ not being available 
experimentally (see the Appendix for further details). 

A large part ($\sim 20$~\%) of the superfluid weight in the underdoped regime 
thus builds up at the expense of spectral weight coming from high energy regions 
of the optical spectrum ($\hbar \omega \geq 1$~eV). Because of our error bars, 
we cannot make a similar statement for the optimally doped and overdoped 
samples, where the sum rule may be exhausted at roughly $500-1000$~cm$^{-1}$. 
The FGT sum rule has been also worked out for the thick Bi-2212 sample using the 
conductivities deduced from the fit. We do find in this case as well that the 
sum rule is satisfied for a conventional energy scale (1000~cm$^{-1}$).

Our results for the optimally doped and overdoped samples thus do not contradict 
earlier similar work in YBCO and Tl$_{2}$Ba$_{2}$CuO$_{6+\delta}$ 
(Tl-2212).\cite{BasovVDM} A recent work on underdoped YBCO shows that, in the 
spectral region where the temperature dependence of the spectra was measured (up 
to about $5000$~cm$^{-1}$), the FGT sum rule is not 
saturated.\cite{Homes-SumRules} On the other hand, our results in the underdoped 
regime are in agreement with recent ellipsometric measurements in the visible 
range which shown that in-plane spectral weight is lost in the visible 
range.\cite{Rubhausen-100Delta,vdmScience} However, the latter results did not 
provide direct evidence that this spectral weight is indeed transferred into the 
condensate.

The remarkable behavior of the underdoped sample must be critically examined in 
light of the uncertainties that enter in the determination of the ratio
$\Delta W/W_{s}$. The determination of $\Delta W$ assumes that $W(T_{A})$ is a 
fair estimate of the spectral weight obtained at $T_{B} < T_{c}$, defined as 
$W_{n}(T_{B})$, {\it if the system could be driven normal at that temperature}. 
While this assumption is correct in BCS superconductors, it may no longer be 
valid for high-$T_{c}$ superconductors.\cite{Chakravarty} Hence, our taking the 
normal-state spectral weight $W(T_{A} \geq T_{c})$ instead of $W_{n}(T_{B})$ 
(unknown) may bias the sum rule.

\begin{figure*}
  \begin{center}
    \begin{tabular}{lcr}
      \includegraphics[width=5.5cm]{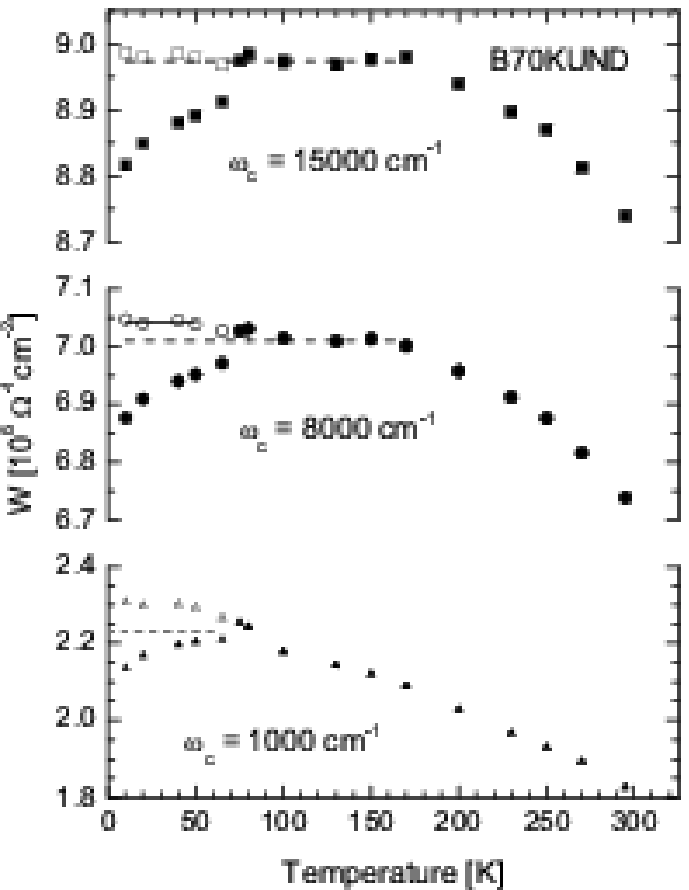} &
      \includegraphics[width=5.5cm]{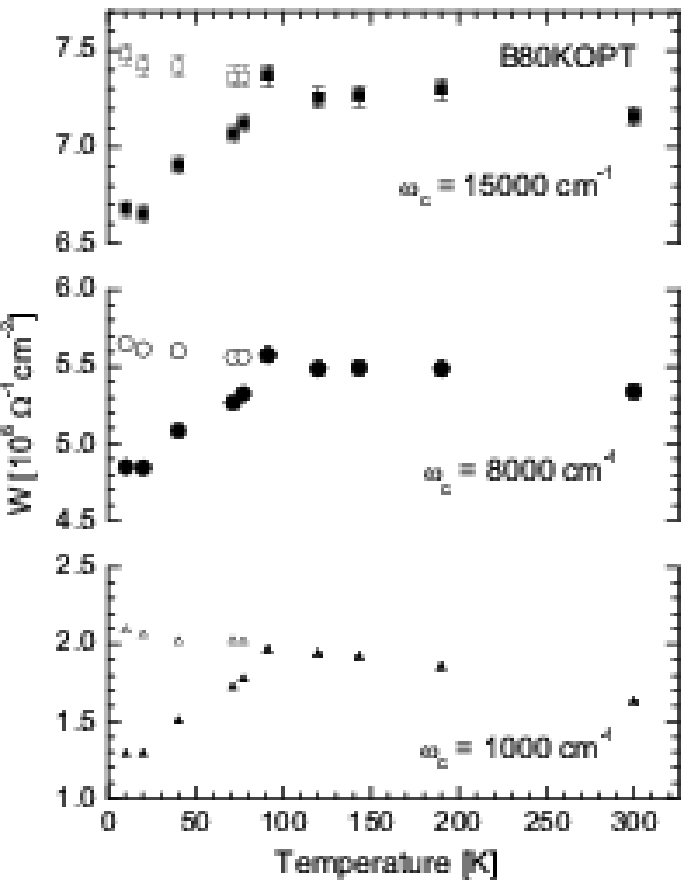} &
      \includegraphics[width=5.5cm]{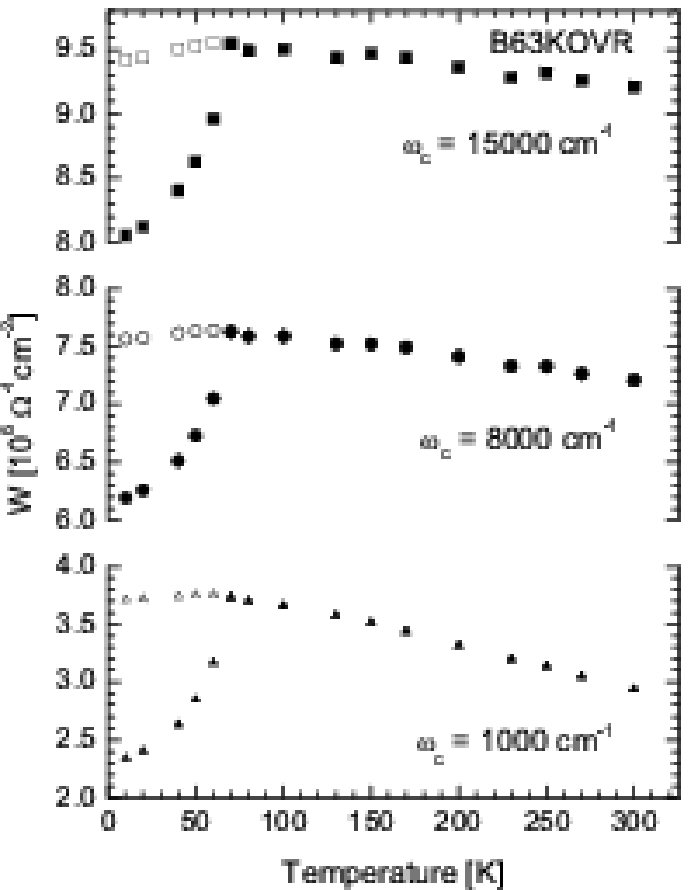}
    \end{tabular}
  \end{center}
  \caption{\label{Fig6}
     Spectral weight $W(T,\omega_c)$
     versus temperature for the underdoped (left), optimally doped (middle)
     and overdoped (right) samples, at different cutoff frequencies $\omega_c$
    (full symbols). $\omega_c=1000$~cm$^{-1}$ (triangles); 8000~cm$^{-1}$
    (circles); and 15000~cm$^{-1}$ (squares). Open symbols
    are obtained by adding the superfluid weight $W_s(T)$ to the spectral
    weight $W(T<T_c)$. In the left panel, the dotted lines show approximately
    the expected location of the open symbols if all the spectral weight
    was removed from low energy states (see text), and the solid line at
    8000~cm$^{-1}$ shows the average value (between 10~K and 50~K) 
    of the open symbols.  In all the curves, the absolute value of the
    spectral weight is subject to a {\it systematic} error of the order of $15-20$\%.
    The errors in the {\it relative} variations of the spectral weight are indicated
    either by the size of the symbols or by the error bars.
    }
\end{figure*}

The error incurred by doing so can be estimated as follows. Figure \ref{Fig6} 
displays the temperature dependence, from 300~K down to 10~K, of the spectral 
weight $W(\omega_c,T)/W(\omega_c,300 {\rm K})$, for three selected integration 
ranges, according to Eq.~\ref{eq:DefW}. At $\omega_c=1000$~cm$^{-1}$, for 
example, the spectral weight exhibits a significant increase as the temperature 
is lowered, and could therefore keep increasing in the superconducting state. 
Hence $W(T_A)$ is most likely to give too small an estimate for $W_n(T_B)$ at 
$T < T_c$, in this energy range.  To get a better insight of this possible 
underestimate, the superfluid weight $W_s(T)$ was added to the spectral weight
$W(T_B)$ (open symbols in Fig.~\ref{Fig6}), at each frequency 
$\omega = \omega_c$ (for temperatures $T < T_c$).  These points represent how 
$W_n(T_B)$ would evolve at $T < T_c$ if {\it all} the superfluid weight were 
accumulated from the spectral range $0^{+} \to \omega_{c}$. One can try to infer 
the evolution of $W_n(T_B)$ from the temperature behavior of the data points 
close to, but still above, $T_{c}$.  For example, for the underdoped sample, at 
$\omega_{c}=1000$~cm$^{-1}$, one could infer an increase of $\sim 4$\% of 
$W_n(T_B)$ from $T_{c}$ down to 10~K. This translates into an absolute error of 
about 0.3 in the value of the FGT sum rule at this frequency.

Such estimates have been performed for a number of cut-off frequencies starting 
from 100~cm$^{-1}$. The error in the FGT sum rule is the largest and most 
ill-defined at low frequencies (where only a rough estimate is possible), but 
becomes negligible at 5000~cm$^{-1}$ and above, where the changes with 
temperature of the normal-state spectral weight should be approximately 10 times 
smaller than at 1000~cm$^{-1}$ and present a smoother behavior when crossing
$T_{c}$. At 15000~cm$^{-1}$ and beyond (Fig.~\ref{Fig6}), the normal-state 
spectral weight is constant for $T<170$~K, meaning that the redistribution of 
spectral weight in the normal state lies within this range of frequencies. Above 
5000~cm$^{-1}$, the uncertainties that we have to deal with are those due to the 
error in the relative change of the measured reflectivity with temperature, to 
the fitting accuracy, and to the determination of $W_s$. The latter two 
uncertainties are not independent and must be calculated self-consistently. 
They yield an upper bound of 15~\%--20~\% in the uncertainty on the evaluation 
of $\Delta W / W_s$, for all frequencies. {\it All} uncertainties are then
represented by the error bars in Fig.~\ref{Fig5}.

>From figure~\ref{Fig6} it is clear as well that any other choice of a 
normal-state temperature $T_{A} \geq T_{c}$ can only increase the violation of 
the FGT sum rule.  

Another point that might be considered is whether, and how, the possible 
existence of a very low frequency mode, below the experimental window of the 
infrared spectroscopy, could give rise to an effect on the FGT sum rule as the 
one observed in this work.  If such mode exists and could not be unveiled by the 
data analysis, then it would affect the evaluation of the normal and 
superconducting spectral weights, and/or of the superfluid density. Assuming 
that such mode has a simple Lorentzian form, one can straightforwardly show that 
its effect on the evaluation of the violation of the FGT rule is {\it at worst} 
$\sim \Delta \epsilon \times (\Omega_{\textsc{t}}/\Omega_{\textsc{l}})^{2}$, 
where $\Delta \epsilon$ is the oscillator strength of the mode, 
$\Omega_{\textsc{t}}$ its frequency, and $\Omega_{\textsc{l}}$ the London 
frequency of the system. Taking $\Omega_{\textsc{t}} \lesssim 10$~cm$^{-1}$ 
(our measurements begin at 30~cm$^{-1}$), and 
$\Omega_{\textsc{l}} \approx 2500$~cm$^{-1}$ (for our underdoped sample), one 
obtains that an effect of the order of 20\% on the FGT sum rule (our 
experimental observation) would need a mode with an oscillator strength 
$\Delta \epsilon \gtrsim 10^{4}$.  If such gigantic oscillator strengths can be 
observed in insulating materials,\cite{Blumberg-BB} they are physically
unreasonable in metallic systems where the low-frequency reflectivity is nearly 
unity.  We have additionally performed numerical simulations on our data to 
evaluate more accurately the possible effect of such modes.  The result is that, 
even if a very-low frequency mode with $\Delta \epsilon \sim 10^{4}$ is included 
in the fittings, but the accuracy of the fit in the measured experimental range 
is kept constant, the associated effect in the FGT sum rule is at most of the 
order of 10\% at frequencies up to $\sim 5000$~cm$^{-1}$, and negligible or
slightly {\it negative} (the sum rule violation is {\it enhanced}) beyond.

For the underdoped sample, the violation of the sum rule, with 
$\Delta W/W_{s}= 0.80 \pm 0.15$ at 8000 cm$^{-1}$, is then clearly established. 
Within the error bars, the sum rule is exhausted in this sample above 
16000 cm$^{-1}$. The fact that the superfluid involves high energy states is 
compatible with the plot below $T_c$ of the sum of the spectral weight at finite 
frequency and the superfluid weight (open symbols in Fig.~\ref{Fig6}). Unlike
the overdoped sample, where at 1000 cm$^{-1}$ the spectral weight of the 
condensate already balances the spectral weight lost up to this frequency, in 
the UD sample the superfluid spectral weight exceeds the loss at 1000~cm$^{-1}$ 
(by roughly a factor of 2). This corresponds to $\Delta W / W_s$ being of order 
50~\% at this energy, consistent with the data of Fig. ~\ref{Fig5} including the 
error bars. At 8000 cm$^{-1}$, for the UND sample, the open symbols of 
Fig.~\ref{Fig6} are consistently above the normal-state spectral weight, 
suggesting that there is still some spectral weight coming from higher energy, 
and consistent with the value for the violation of the sum rule at this 
frequency.

\subsection{Violation of the sum-rule, change of kinetic energy and
            pairing mechanism}

One interpretation of the sum rule violation can be made in the context of the 
tight-binding Hubbard model. The relation between the spectral weight over the 
conduction band and the kinetic energy $E_{kin}$ per copper site 
yields:\cite{vdMTrieste}
\begin{equation}
  \frac{\Delta W}{W_s}+\frac{4\pi c}{137\hbar}\frac{a^2}{V}
  \frac{1}{\Omega_L^2} (E_{kin,n}-E_{kin,s}) = 1
 \label{eq:FGTEk}
\end{equation}
where $a$ is the (average) lattice spacing in the plane and $V$ is the volume 
per site (SI units). This relation means that a breakdown of the FGT sum-rule up 
to an energy $\hbar \omega_c$ of the order of the plasma frequency ($\sim 1$~eV 
for Bi-2212) is related to a change in the carrier kinetic energy 
$\Delta E_k = E_{kin,n}-E_{kin,s}$, when entering the superconducting state.
According to our results in the underdoped sample (Fig.~\ref{Fig5}), 
$\Delta W/W_s=0.80 \pm 0.15$ at 1 eV and 
$\Omega_L^{2}=6.52 \times 10^6$~cm$^{-2}$, which yields 
$\Delta E_{k} =(0.85 \pm 0.2)$~meV per copper site. This would be a huge kinetic 
energy gain, $\sim 10$ times larger than the condensation energy $U_0$. For 
optimally doped Bi-2212, $U_0 \simeq 1$~J$/$g-at~$\approx 0.08$~meV per copper 
site.\cite{Loram}

A similar, though more robust estimate of the kinetic energy gain for the 
underdoped sample can be made from figure~\ref{Fig6} by taking into account the 
{\it temperature evolution} of the spectral weight in the normal and 
superconducting states, and not just the difference between the spectral weights 
at 100~K and 10~K. At 8000~cm$^{-1}$, when the superfluid weights are included 
below $T_{c}$ (open symbols), figure~\ref{Fig6} represents the intraband 
spectral weight, hence $-E_k$, as a function of temperature. Below 170~K and 
above the resistive transition at 100~K, the normal-state spectral weight for 
the underdoped sample levels-off (its average value is represented by the dotted 
line), and we observe a clear excess of spectral weight in the superconducting 
state with respect to the average normal-state spectral weight. Taking as well 
the average of the low-temperature superconducting spectral weight (solid line
through the open symbols) we find $\Delta E_k = (0.83 \pm 0.2)$~meV per copper 
site, in agreement with our estimate from the violation of the FGT sum rule. If 
we incorporate the two data points at 80~K and 75~K (within the resistive 
transition), we find $\Delta E_k = (0.5 \pm 0.2)$~meV per copper 
site.\cite{Andres-Miami}

Various scenarios are consistent with our results. Most of them (but not all) 
propose indeed  a superconducting transition driven by an in-plane kinetic 
energy change: holes moving in an antiferromagnetic 
background,\cite{AFEk-1,AFEk-2,AFEk-3} hole 
undressing,\cite{Hirsch1-1,Hirsch1-2,Hirsch2} phase fluctuations of the 
superconducting order parameter in superconductors with low carrier density 
(phase stiffness),\cite{Eckl-PhaseFluctDEk} and quasiparticle formation in the 
superconducting state.\cite{NormanPepin-QPFormation} There is a quantitative
agreement between our results and those of the hole undressing and quasiparticle 
formation scenarios. Other models have been recently discussed in order to 
account for these data.\cite{Benfatto-Model,Stanescu-Model,Ashkenazi-Model,
Carbotte-Model,Wrobel-Model,Kopec-Model}

Recently, STM experiments in optimally doped Bi-2212 samples showed small scale 
spatial inhomogeneities, over $\simeq 14$ {\AA}, which are reduced significantly 
when doping increases, and whose origin could be local variations of oxygen 
concentration.\cite{Davis} Since the wavelength in the full spectral range is 
larger than 14 \AA, the reflectivity performs a large scale average of such an 
inhomogeneous medium. The implications in the conductivity are still to be 
investigated in detail, but it is presently unclear how this could affect the 
sum rule.

%
\begin{figure*}
  \begin{center}
    \begin{tabular}{c||c}
      \includegraphics[height=7.5cm]{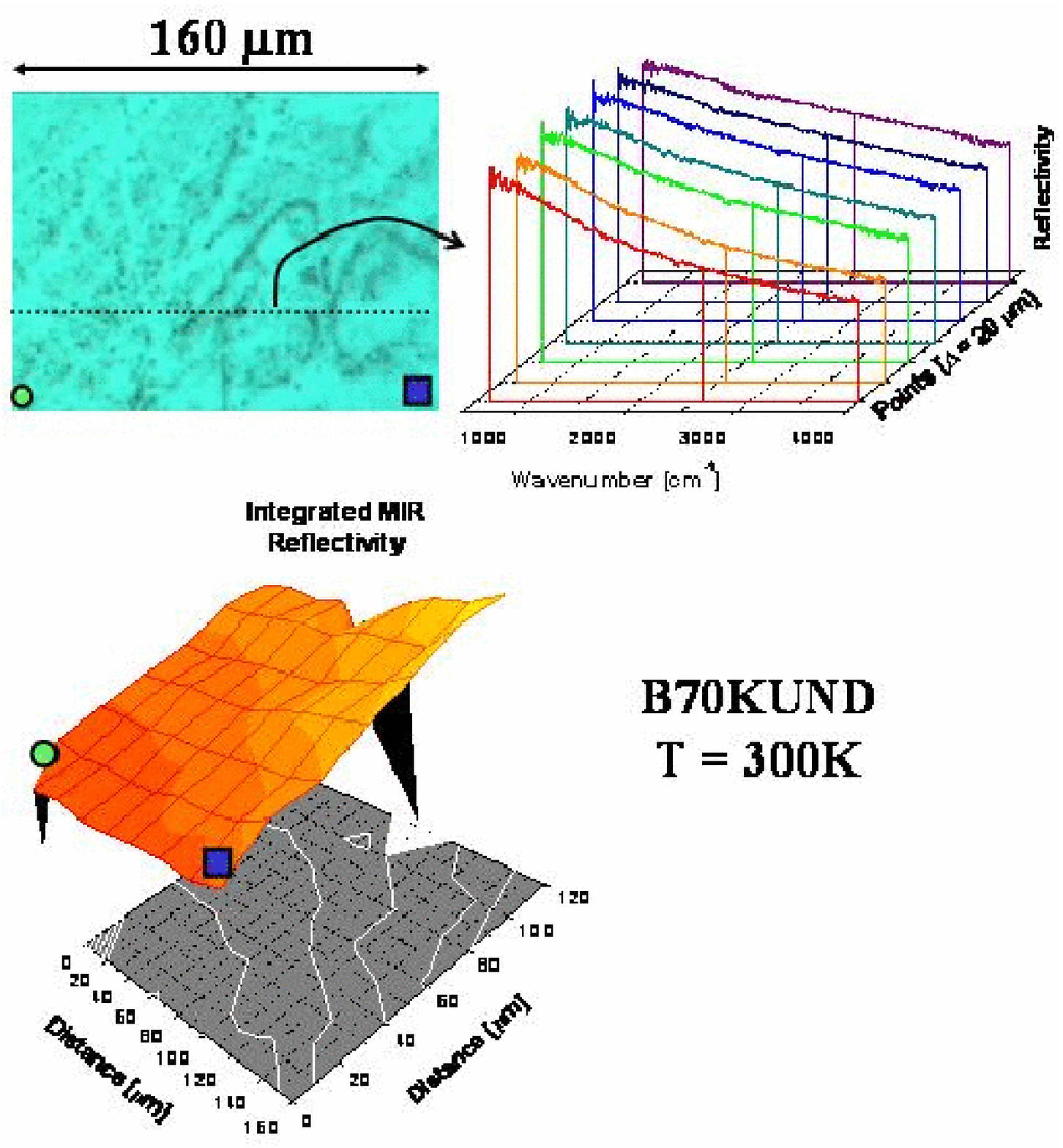} & 
      \includegraphics[height=7.5cm]{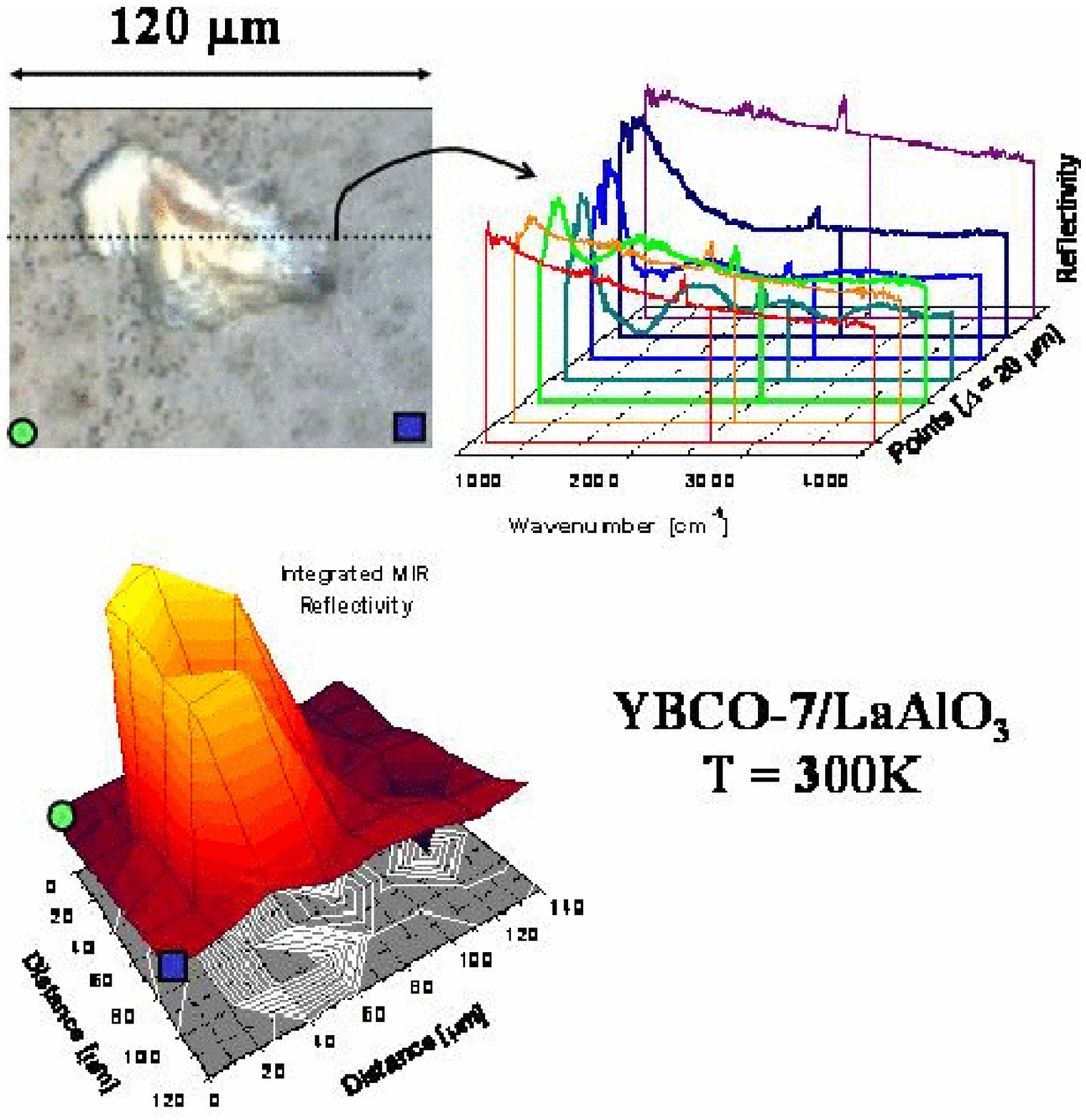}   
    \end{tabular}
  \end{center}
  \caption {\label{Fig7}
    Left: infrared chart of the B70KUND sample. The picture of the
    scanned region (upper left corner) shows some sub-micronic defects. The
    reflectivity spectra along the dotted line (upper right corner) are
    identical, and the integrated reflectivity chart (lower left corner)
    is homogeneous. (The two peaks correspond to mechanical instabilities
    in the microscope sample holder during the measurements, and can not
    be reproduced). Right: infrared chart of a YBCO-7 sample, having surface
    defects larger than 10~$\mu$m.  Note that the reflectivity spectra along
    the dotted line, and the integrated reflectivity chart, both show
    accidents clearly correlated with the surface defect seen on the picture. }
\end{figure*}


\section{CONCLUSIONS}
\label{sec:conclusions}

In conclusion, we have found for the in-plane conductivity of the underdoped 
Bi-2212 a clear violation of the FGT sum rule at 1~eV, corresponding to a large 
($\sim 20$~\%) transfer of spectral weight from regions of the visible spectrum 
to the superfluid condensate. Within the framework of the tight-binding Hubbard 
model, this corresponds to a kinetic energy lowering of $\sim 0.5-1$~meV per
copper site. The very large energy scale required in order to exhaust the sum 
rule in the underdoped sample cannot be related to a conventional bosonic scale, 
hence strongly suggests an electronic pairing mechanism.


\acknowledgments
We are very grateful to M. Norman, C. P\'epin and A.~J.~Millis for illuminating
discussions. We acknowledge fruitful comments from J. Hirsch and E.Ya. Sherman. 
We thank P. Dumas (LURE, Orsay) for his help with the IR microscopy 
measurements and A. Dubon and J.Y. Laval (ESPCI) for the electron microscopy of 
the UD sample. AFSS thanks Colciencias and Minist\`ere Fran\c cais des Affaires 
Etrang\`eres (through the Eiffel Fellowships Program) for financial support.
This work was partially supported by ECOS-Nord-ICFES-COLCIENCIAS-ICETEX through 
the ``Optical and magnetic properties of superconducting oxides" project, 
Universidad de Antioquia and through the COLCIENCIAS project 1115010115
``Estudio de propriedades \'opticas de superconductores de alta temperatura 
cr\'itica".

\section{APPENDIX}
\label{sec:appendix}


\subsection{Infrared microscopy characterizations}

Even if infrared and visible spectroscopy measures the bulk optical properties 
of cuprates (the infrared skin-depth of Bi-2212 is $\gtrsim 3000$~\AA), a flat 
surface of high optical quality {\it in the whole measured spectral range} is an 
important requirement for the accuracy in the measurement of the {\it absolute} 
value of the reflectivity. Besides checking our samples under an optical 
microscope, we characterized the {\it mid-infrared} optical quality of our films 
by infrared microscopy at room temperature, using the infrared microscope of the 
MIRAGE beamline at LURE (Orsay).\cite{MIRAGE} In this microscope, the light 
being focussed on a point on the sample has been previously analyzed by a 
Fourier Transform spectrometer. In this way, the infrared reflectivity spectrum 
for each measured point is available. In our case, we worked at a spatial 
resolution of 20~$\mu$m. We then made, for each sample, scans over zones 
typically $160 \times 160$~$\mu$m$^{2}$, recording for each measured point the 
spectrum in the $[500-4000]$~cm$^{-1}$ range.

Figure~\ref{Fig7} illustrates  the cases of a good (left) and a bad (right) 
optical quality in the mid-infrared [the figure in the right corresponds 
actually to the  YBCO thin film, with a $T_{c}=90$~K]. We found, not 
surprisingly, that a defect whose size is comparable to the infrared wavelength 
($\gtrsim 10$~$\mu$m), and that is easily spotted with the visible microscope 
(picture in Fig.~\ref{Fig7}, right), can perturb drastically the local 
reflectivity spectra. (We have not, on the other hand, found any case where a 
smooth region in the optical microscope was defective under the infrared 
microscope). If the ensemble of these infrared-defective regions represents a 
few percent of the total sample surface, then the {\it average} reflectivity of 
the whole surface will not accurately represent the {\it intrinsic} value of the 
absolute reflectivity of the sample ({\it i.e.}, the average reflectivity over 
the whole surface will differ by more than 1\% from the actual value of the 
reflectivity one would obtain with a sample of the same doping characteristics 
but with a surface free of defects). Our characterizations show as well that the 
infrared spectra are not sensitive to sub-micronic defects, whose size is 
comparable to the wavelengths in the visible (Fig.~\ref{Fig7}, left).

\subsection{Sample selection}

A first screening of the samples was done based on their resistive, 
optical-microscopy and $\mu$IR characteristics. We selected the samples showing 
a metallic behavior in the normal state, having the narrowest transitions 
($\sim 20$~K to $\sim 10$~K for UND to OVR, respectively), homogeneous under the 
optical microscope, and displaying homogeneous $\mu$IR charts. The reflectivity 
spectra of the samples thus selected were adjusted to obtain their 
conductivities, using a fitting procedure that will be discussed in the next 
section. In this fitting procedure, the thickness of the sample enters as a 
parameter. Samples whose measured thickness differed by more than 15\% (the 
accuracy of RBS measurements on Bi-2212) from the thickness needed to adjust the 
measured reflectivity were discarded. Samples showing abnormal electronic
behavior ({\it e.g.}, localized states at low frequency, or a temperature 
dependence of the low-frequency conductivity not matching the DC measurements) 
were discarded as well. At the end, three samples having passed confidently all 
the screenings, and spanning the three different doping regions, were selected 
for a thorough analysis. The critical temperatures (doping) of the selected 
films are 70~K (UND), 80~K (OPT) and 63~K (OVR).

The thickness of the films is an important issue: it is needed to deduce the 
intrinsic optical functions of the Bi-2212 despite the presence of a substrate 
which modifies the overall reflectivity. After the optical measurements were 
completed, the thickness of the B80KOPT and B63KOVR samples was directly 
verified by a second run of RBS measurements , yielding 395~nm and 270~nm 
($\pm 30$~nm) respectively, in agreement with the thicknesses derived from 
fitting their reflectivities. The uncertainty of the RBS measurement (larger 
than 10\%) is not quite satisfactory, since as discussed further, it is possible 
to change the assumed thickness within 10\% and the incurred error bar is 
$\sim 10$\%. The B70KUND sample was crucial, and the RBS measurement on a film 
grown under the same conditions (but on a MgO substrate) yielded 220~nm, already
in agreement (within 10\%) with the thickness deduced from the fit. We therefore 
decided to try and achieve a better accuracy using a TEM measurement, thus 
precluding a second RBS run. This attempt was unsuccessful.


\subsection{Thin-films: extraction of the Bi-2212 optical functions}

Among the various different models that we tried for a film on top of a 
substrate, we found that the best description of our raw reflectivity spectra is 
obtained by assuming that the light bouncing back and forth within the film 
interferes coherently and that the substrate has an infinite thickness (no light 
is reflected from its rear face).  The typical thickness of our films being 
3000~{\AA}, that of the substrate $\sim 0.5$~mm, the refractive index of the 
Bi-2212 film (F) in the mid-infrared (around 800~cm$^{-1}$, for example) being 
$\eta_{\textsc{f}} \approx 5$ and that of the substrate (S) 
$\eta_{\textsc{s}} \sim 0.1 - 5$, one can deduce that, in this spectral range, 
the ratio ``optical path~$/$~wavelength in the medium" for a round-trip path of 
the light is about 0.24 for the Bi-2212, and $8-40$ for the substrate. On the 
other hand, even if the substrate is transparent in the near-infrared and the 
visible, its rear face in our films is not polished, so that the light is 
scattered from this face. The above-mentioned assumptions are thus largely 
justified on physical grounds.

The interfaces playing a role in our model are thus helium bath-film (HF) and 
film-substrate (FS).  They are characterized by complex reflection 
[$r_{12}=(n_{1}-n_{2})/(n_{1}+n_{2})$] and transmission 
[$t=2n_{1}/(n_{1}+n_{2})$] coefficients from medium 1 (refractive index $n_{1}$) 
to medium 2 (refractive index $n_{2}$). The refractive index of the helium bath 
around the sample is 1. Denoting by $\rho$ the complex reflection coefficient of 
our films (the measured reflectivity being thus $R=|\rho|^{2}$), and by $\delta$ 
the thickness of the film, and assuming a coherent interference within the film 
and an infinite substrate, one easily finds
\begin{equation}
  \rho = r_{\textsc{hf}} +
  \frac{t_{\textsc{hf}}t_{\textsc{fh}}r_{\textsc{fs}}A_{\textsc{f}}}
  {1-r_{\textsc{fs}}r_{\textsc{fh}}A_{\textsc{f}}},
\label{eq:reflec-thin-film}
\end{equation}
where
$A_{\textsc{F}}(\omega,\delta)=\exp[2i(\omega/c)\eta_{\textsc{f}}(\omega)\delta]$.
Therefore, when the complex refractive indices of the film and the substrate 
(entering into the various coefficients $r$ and $t$) and the thickness of the 
film are known, one can describe the measured reflectivity $R$.

We have measured the reflectivity of a single crystal of SrTiO$_{3}$ for all the 
temperatures and spectral range relevant to our work.  We have thus at hand the 
optical functions of the substrate. Hence, the measured reflectivity spectra can 
be adjusted by means of the Eq.~\ref{eq:reflec-thin-film} by simulating only the 
refractive index of the Bi-2212.  The choice of the essay functions to simulate 
$\eta_{\textsc{f}}$ is {\it arbitrary}, provided the chosen functions obey 
{\it causality} (see next section).  When the fitting is exact, this procedure 
yields the exact optical functions of the film alone, and the film thickness.  
The latter has been controlled by RBS measurements (see main text). The 
uncertainties arising from the accuracy of the fit are discussed in the next 
section.


\subsection{Thin-film reflectivity: analysis of uncertainties and errors}

If ${R(\omega), \omega \in \left[ 0,\infty \right)}$ is the {\it complete} and 
{\it exact} reflectivity spectrum of a material, then the phase $\theta(\omega)$ 
of the complex reflectance coefficient of the system can be deduced from a 
Kramers-Kronig transformation (TKK) on $R(\omega)$ alone:\cite{Wooten}
\begin{equation}
  \theta (\omega_{0}) = -\frac{\omega_{0}}{\pi} \int_{0}^{\infty}
  \frac{\ln [R(\omega)/R(\omega_{0})]}{\omega^{2}-\omega_{0}^{2}} d\omega,
\label{eq:TKK}
\end{equation}
This relation arises solely from the assumption of a linear and causal response 
of the system to the electromagnetic excitation.

The knowledge of $R$ and $\theta$ allows the determination of any optical 
function $f(\omega)$ by means of simple algebraic relations. We will write
\begin{equation}
  f(\omega)=f[R(\omega),\theta(\omega)].
\label{eq:OptFn-Gral}
\end{equation}

It is evident, from~(\ref{eq:TKK}) and~(\ref{eq:OptFn-Gral}), that if the 
reflectivity $R(\omega)$ of a given system is known, and if we can exactly fit 
this reflectivity with {\it any arbitrary} essay function {\it obeying 
causality}, then the optical functions issued from this essay function are {\it
mathematically identical} to those issued from a Kramers-Kronig transformation 
of $R(\omega)$.  In other terms, {\it an exact fitting from zero to infinity 
with a causal function is completely equivalent to doing TKK}.

Experimentally, the reflectivity is only known within a spectral range 
$[\omega_{a}, \omega_{b}]$.  Besides, the reflectivity is never measured with 
zero uncertainty, neither in absolute value nor in its relative (frequency- 
and/or temperature-) variations. These two facts affect the accuracy in the 
determination of the optical functions (either in the fitting or TKK cases), and 
are the departure point for the analysis of the uncertainties and errors in the 
determination of the optical functions of a system from its {\it measured} 
reflectivity. In the case of fitting, one additional source of error is the 
accuracy of the fitting within the measured frequency window.

The uncertainty in the determination of an optical function at a given frequency  
depends on the uncertainty on the determination of the reflectivity at {\it all} 
the frequencies, as seen from relation~\ref{eq:TKK}.  Be $\Delta R (\omega)$ and 
$\xi (\omega) \equiv \Delta R(\omega)/R(\omega)$, respectively, the absolute and
relative errors in the determination of the reflectivity. The questions at stake 
in the analysis of the errors in the optical functions are thus:
\begin{enumerate}
  \item Given the maximum value $|\xi|_{\textrm{max}}$ that $\xi
  (\omega)$ can take in a frequency range, what is the maximum error
  expected in the determination of the optical functions?
  \item How, in the calculation of the optical functions from
  the reflectivity, does the error $\xi (\omega)$ propagate to
  frequencies $\omega \prime \neq \omega$?
\end{enumerate}
%
\begin{figure}
  \begin{center}
      \includegraphics[width=7cm]{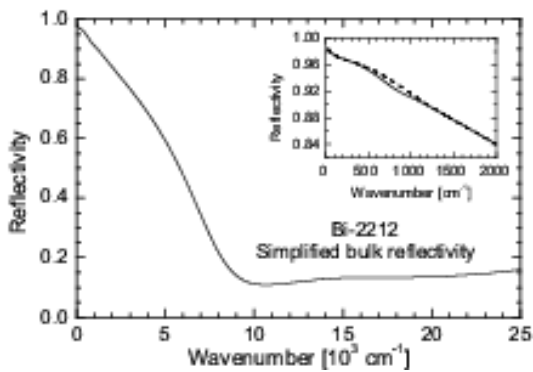}
  \end{center}
  \caption{\label{Fig8}
    Simulated reflectivity for Bi-2212 in the
    normal state, in the spectral range $0-25000$~cm$^{-1}$,
    constructed from a superposition of a
    Drude plus various broad Lorentz oscillators.  The inset
    shows a Gaussian ``perturbation" (dashed spectrum)
    to this ``exact" reflectivity, used as a model
    for a fitting or measure mismatch in the mid-infrared.
    }
\end{figure}

To answer these questions, let us call $\Delta \theta$ the error, due to 
$\Delta R$, in the determination of the phase of the reflectance coefficient. 
Assuming that $|\Delta R|$ and $|\Delta \theta|$ are small compared to unity 
(which is always true in the measured spectral range), then the relative errors 
in the determination of any optical function can be expressed in terms of $R$, 
$\theta$, $\xi$ and $\Delta \theta$ as:
\begin{equation}
  \frac{\Delta f(\omega)}{f(\omega)} =
  C_{\xi}(\omega)\xi(\omega)+ C_{\theta}(\omega)\Delta
  \theta(\omega).
\label{eq:Error-f}
\end{equation}
The dimensionless factors $C_{\xi}$ and $C_{\theta}$ are different for the 
different optical functions of a given system. In particular, for the 
conductivity, it can be shown that $C_{\xi} \sim (1-R)^{-1}$. Besides, they have 
a complicated explicit dependence on $R(\omega)$ and $\theta(\omega)$, so that, 
for a given optical function, their frequency-dependencies are different for 
systems with different reflectivities.  Hence, $\Delta f/f$ depends on the 
explicit form of the reflectivity spectrum, and the examination of the 
uncertainties in the determination of the optical functions has to be made on a 
case-by-case basis. While an analytical approach to the study of these 
uncertainties in the case of cuprates can be performed, the calculations 
involved are too lengthy (though simple; they can be found in 
reference~\cite{Andres-Thesis}) and yield too crude results. For a system with a 
non-trivial reflectivity spectrum like Bi-2212, it is far more reliable to 
perform some numerical simulations and to compare them with the analytical 
studies.

Figure~\ref{Fig8} shows a simplified typical reflectivity for Bi-2212. This 
reflectivity has been constructed by superposing a Drude and various broad 
Lorentz oscillators. It corresponds approximately to the reflectivity of an 
overdoped Bi-2212 sample in the normal state. Analyzing an overdoped-like
reflectivity that is close to unity at low frequencies sets an upper limit, in 
this frequency range, to the errors incurred in the determination of the optical 
functions. Let us then define a ``Gaussian perturbation" 
$\xi_{\textsc{g}}(\omega)$ (center $\omega_{c}$, half-width $\sqrt{2}\sigma$) to 
this exact reflectivity:
\begin{equation}
  \xi_{\textsc{g}}(\omega) = \alpha
  \exp{[-\frac{(\omega-\omega_{c})^{2}}{2\sigma^{2}}]}.
\label{eq:PertGauss}
\end{equation}
This kind of perturbation represents reasonably well the type of errors expected 
when fitting the experimental reflectivity using Drude and broad Lorentz 
oscillators, as we have done in this work. We will take 
$\alpha = |\Delta R/R|_{\textrm{max}}=1$\%, $\omega_{c}=800$~cm$^{-1}$ and 
$\sigma=200$~cm$^{-1}$, as an example of a fitting mismatch of 1\% in the 
mid-infrared (inset of Fig.~\ref{Fig8}).
%
\begin{figure}
  \begin{center}
      \includegraphics[width=7cm]{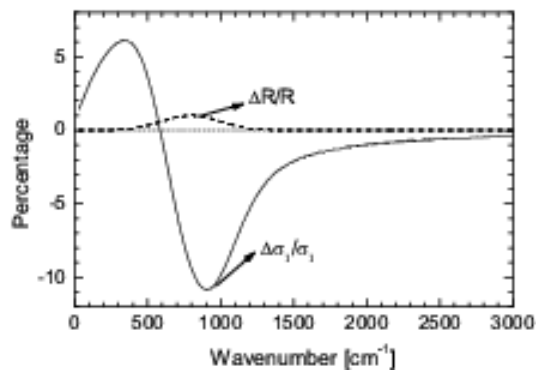}
  \end{center}
  \caption{\label{Fig9}
    Relative uncertainty in the real conductivity (full line)
    introduced by a ``Gaussian perturbation" on the
    mid-infrared reflectivity (dashed line).  Note the change of sign
    of $\Delta \sigma_1 / \sigma_1$, which was eventually neglected
    in the calculation of the error bars shown in the manuscript,
    leading to an overestimate of such error bars.
    }
\end{figure}

Numerical Kramers-Kronig transformations using the same kind of extrapolations 
for both reflectivities can then be performed to obtain the ``exact" and 
``perturbed" conductivities, and then the relative error in the conductivity 
introduced by the perturbation. This relative error is shown in Fig.~\ref{Fig9}, 
where it is clear that 
$|\Delta \sigma_{1}/\sigma_{1}|_{\textrm{max}} \sim 5-10 |\Delta R/R|_{\textrm{max}}$, 
and that the perturbation in the conductivity extends over a region of about 3 
times that of $\Delta R/R$.

The latter example illustrates well the uncertainties endemic of a fitting 
procedure.  Two other kinds of uncertainties play a role in the determination of 
the optical functions from reflectivity data, and are to be considered whichever 
method (TKK or fitting) is used for analyzing the data. These are: the error in 
the absolute measurement of the reflectivity, and the effect on the measured
window of the (unknown) low- and high-frequency response of the system (the 
problem of the choice of extrapolations when doing TKK).

Analytical studies and numerical simulations~\cite{Andres-Thesis} show that, for 
a system like Bi-2212, and with an accuracy in the absolute measurement of the 
reflectivity of the order of 1\% (our case), the associated uncertainties in the 
conductivity become important (20\% or more) at frequencies below 
$\sim 100$~cm$^{-1}$.  On the other hand, as this is a systematic error, it can 
be largely suppressed by comparing two conductivity curves of the same sample. 
This is precisely what is done in the evaluation of the FGT sum rule or in the 
calculation of the relative changes of spectral weight with temperature. 
Besides, the comparison between the extrapolated zero-frequency conductivity and 
the resistivity (Fig.~\ref{Fig4}) narrows the confidence range of the 
low-frequency conductivity.

As for the effects of the low- and high-frequency extrapolations, the conclusion 
of our analysis and simulations is that they can perturb the determination of 
the optical functions up to (at most) twice the first measured frequency and 
down to (at most) $1/3$ of the last measured frequency.


\end{document}